\documentstyle[12pt,epsf,epsfig,psfig]{article}
\def\be{\begin{equation}}
\def\ee{\end{equation}}

\def\lsim{\raise0.3ex\hbox{$<$\kern-0.75em\raise-1.1ex\hbox{$\sim$}}}
\def\gsim{\raise0.3ex\hbox{$>$\kern-0.75em\raise-1.1ex\hbox{$\sim$}}}

\def\NP{{ Nucl.\ Phys.\ }}
\def\PL{{ Phys.\ Lett.\ }}
\def\PR{{ Phys.\ Rev.\ }}

\begin{document}

\thispagestyle{empty}

\noindent September 22, 2001 \hfill BI-TP 2001/20

\vskip 1.5 cm

\centerline{\large{\bf Cluster Percolation and}}

\medskip

\centerline{{\large{\bf Thermal Critical Behaviour}}\footnote{Plenary 
talk given at the {\sl Europhysics Conference on Computational Physics},
Aachen/Germany, Sept. 5 - 8, 2001.}}

\vskip 0.5cm

\centerline{\bf Helmut Satz}

\bigskip

\centerline{Fakult\"at f\"ur Physik, Universit\"at Bielefeld}
\par
\centerline{D-33501 Bielefeld, Germany}

\vskip 0.5cm

\noindent

\centerline{\bf Abstract:}

\medskip

Continuous phase transitions in spin systems can be formulated as
percolation of suitably defined clusters. We review this equivalence and
then discuss how in a similar way, the color deconfinement transition in
SU(2) gauge theory can be treated as a percolation phenomenon. In the
presence of an external field, spin systems cease to show thermal
critical behavior, but the geometric percolation transition persists
(Kert\'esz line). For $H\not=0$, we study the relation between
percolation and pseudocritical behavior, both for continuous and first
order transitions, and show that it leads to the necessity of an
$H$-dependent cluster definition. A viable formulation of this kind
could serve as definition of deconfinement in QCD with dynamical quarks.

\vskip 0.5cm

\noindent{\bf 1.\ Introduction}

\medskip

Critical phenomena in spin systems and elsewhere are usually described
in terms of spontaneous symmetry breaking or of singular behavior of the
partition function. On the other hand, the size of domains of parallel
spins generally increases near criticality, and so the search for a
geometric ``connectivity" formulation has been a long standing
challenge. For some classes of systems, the problem is solved, and I
begin by recalling how this is done.

The Ising model is defined by the Hamiltonian
\be
{\cal H}(T,H) = - J \sum_{<ij>} s_i s_j - H \sum_i s_i;~~~ s_i=\pm 1
\label{1.1}
\ee
which is used to calculate the partition function
\be
Z(T,H) = \sum_{\{s_i\}} \exp\{-\beta {\cal H}(T,H)\}; ~~~\beta\equiv
1/T;
\label{1.2}
\ee
here $J$ denotes the nearest neighbor spin-spin coupling and $H$ an
external field. Given $Z(T,H)$, the magnetization and (isothermal)
susceptibility are obtained as
\be
m(T,H) = (\partial \log Z / \partial H)_T, ~~~
\chi(T,H) = (\partial m / \partial H)_T.
\label{1.3}
\ee
For vanishing external field, $H=0$, the Ising model leads to a critical
point $T=T_c$: for $T < T_c$, the $Z_2$ symmetry of the Hamiltonian
${\cal H}(T,0)$ is spontaneously broken by the state of the system. The
magnetization becomes the order parameter for the transition and
vanishes for $T \to T_c$, while the susceptibility diverges there:
\be
m(T,0) \sim (T_c - T)^{\beta}, ~~~ T \leq T_c; ~~~
\chi(T,0) \sim |T_c - T|^{\gamma}.
\label{1.4}
\ee
The critical exponents $\beta$ and $\gamma$ specify the critical
behavior and define the universality class of the transition.

It was shown by Fortuin and Kasteleyn \cite{FK} that the partition
function of the model, given in Eq.\ (\ref{1.2}) as sum over spin
configurations, can for $H=0$ be equivalently written as a sum over
configurations of clusters containing $n$ lattice points. In the
resulting form
\be
Z(T,0) = \sum_n \{ \prod_{<ij>}^{n_{ij}=1} p_b
\prod_{<ij>}^{n_{ij}=0} (1 - p_b)~ 2^{c(n)} \},
\label{1.5}
\ee
$p_b \equiv 1 - \exp\{-2 \beta J\}$ defines a bond weight between
adjacent sites, while $n_{ij}=1(0)$ specifies whether two sites are
bonded or not. The number of $n$-clusters is given by $c(n)$.

In percolation theory, the observable corresponding to the order
parameter is the percolation strength $P(T)$, defined as the probability
that a given site belongs to a percolating cluster. In place of the
susceptibility we now have the average cluster size $S(T)$; in the
percolation region only finite clusters are included in the averaging.
Using the FK formalism, Coniglio and Klein \cite {CK} showed that the
percolation point coincides with the thermal critical temperature $T_c$,
and that the percolation behavior is governed there by
\be
P(T) \sim (T_c - T)^{\beta},~ T \leq T_c;  ~~~ S(T) \sim
|T_c -T|^{\gamma},
\label{1.6}
\ee
where $\beta$ and $\gamma$ are the usual thermal Ising exponents. For
the Ising model, thermal critical behavior can thus be equivalently 
described as percolation of FK clusters.

\begin{figure}[htb]
\mbox{
\hskip0.5cm\epsfig{file=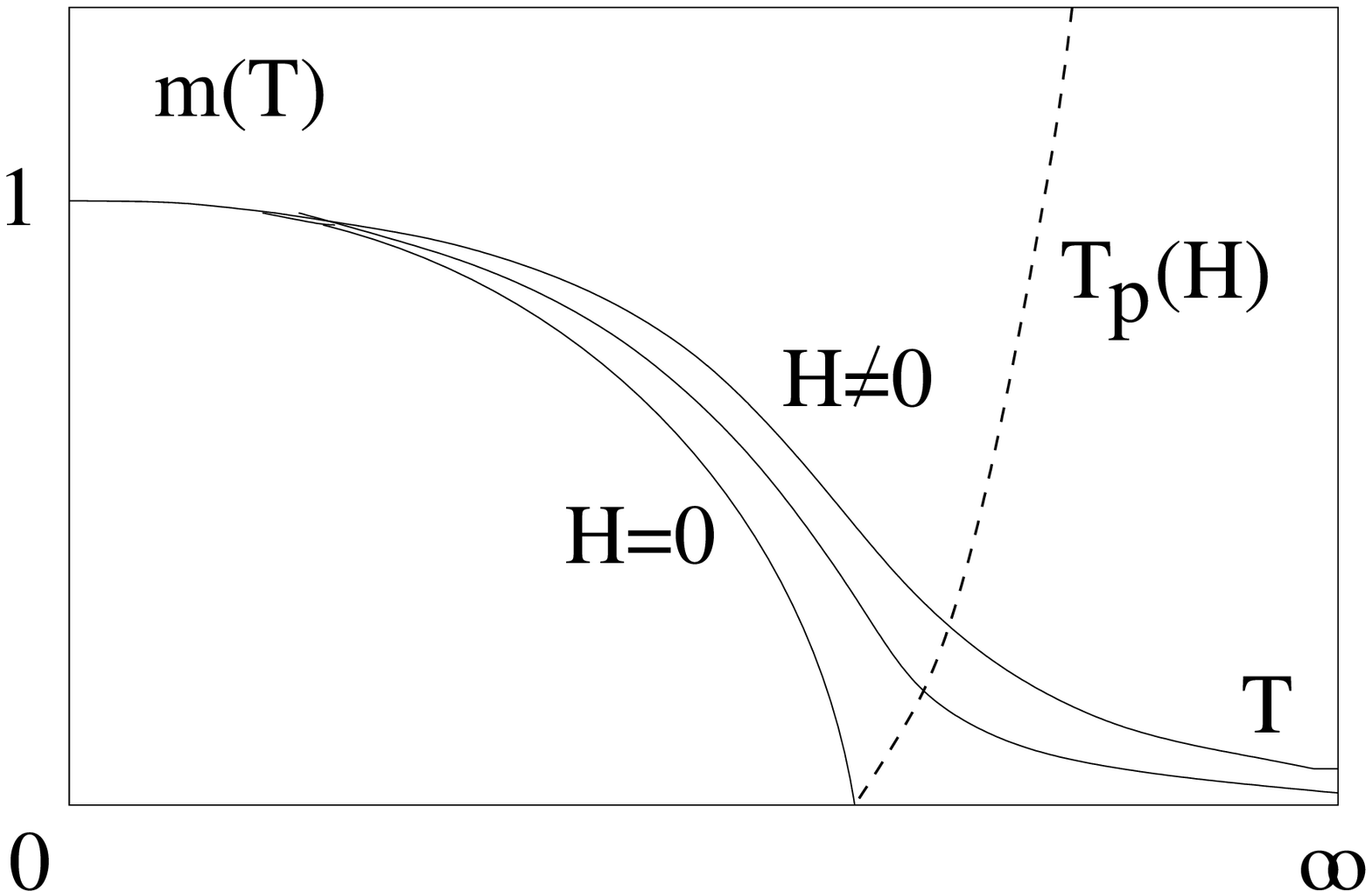,width=6cm,height=4.5cm}
\hskip2cm
\epsfig{file=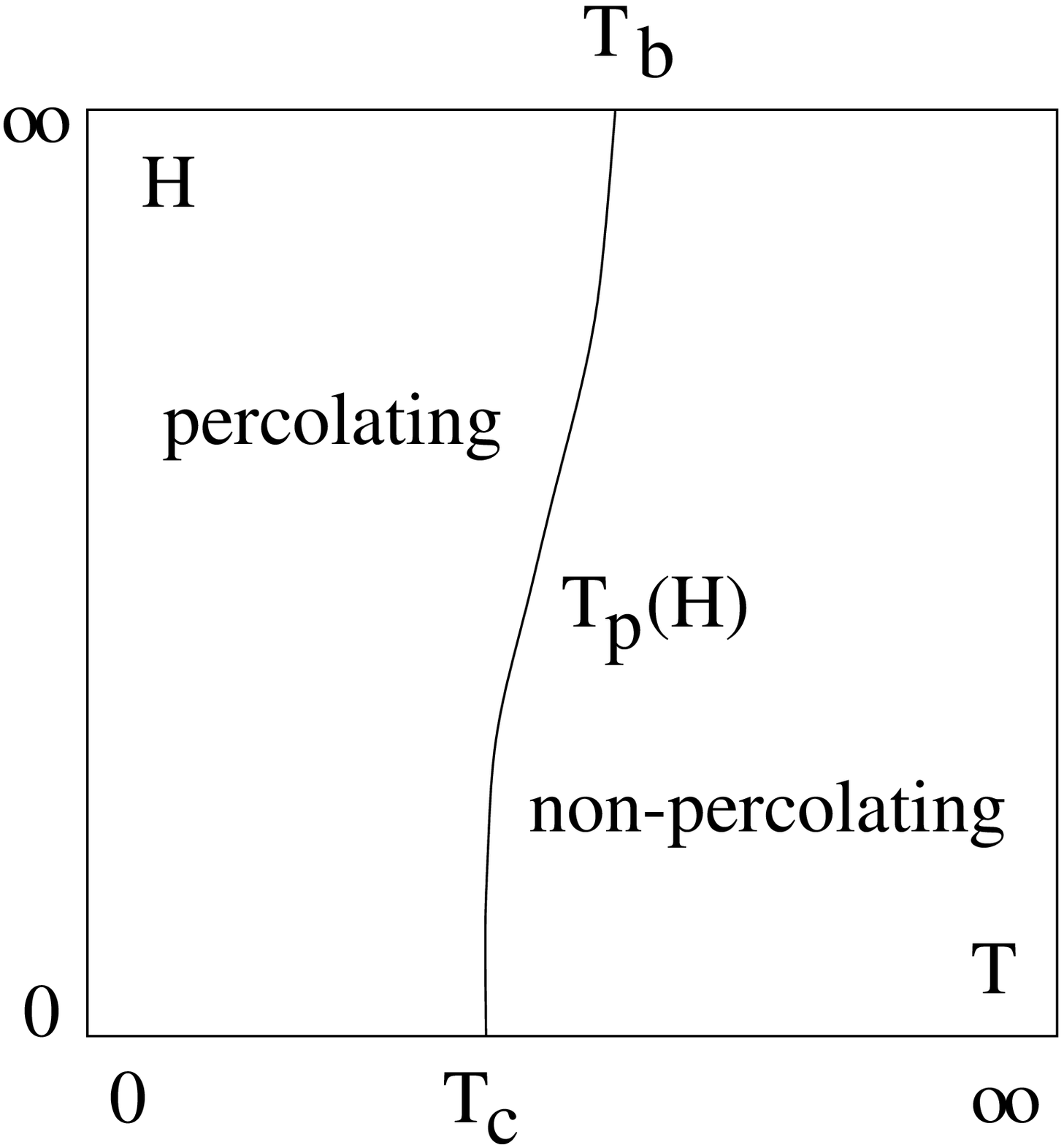,width=6cm, height=5cm}}
\vskip0.5cm
\caption{Magnetization and Kert\'esz line in the Ising model.}
\label{1-1}
\end{figure}

A non-vanishing external field, $H \not= 0$, explicitly breaks the
previous $Z_2$ symmetry of the Hamiltonian, and as a result, partition
function and thermodynamic observables become analytic: there is no more
thermal critical behavior. The magnetization now never vanishes;
$H$ aligns some spins at any temperature (see Fig.\ \ref{1-1}).

However, the percolation transition persists also for $H\not= 0$. For
any value of $H$ there exists a percolation point $T_p(H)$, such that
for $T \leq T_p$ there is percolation and for $T > T_p$ not \cite{K}.
The critical behavior along the ``Kert\'esz line" $T_p(H)$ (see Fig.\
\ref{1-1}) is now determined by percolation critical exponents, so that the
transition is no longer in the Ising universality class.

We note here that the equivalence between thermal and percolation
behavior can in fact be restored if clusters are not required to be
spatially connected. The introduction of an $H$-dependent ghost spin
not on the lattice but able to couple to any lattice site allows
spatially disjoint sites to form a cluster \cite{SW} and thus leads to
percolation for all $H\not= 0$. We are here interested in the relation
between spatial connectivity and thermal behavior; therefore we shall
not pursue the ghost spin approach any further.

One major reason for our study is to address critical behavior in finite
temperature QCD; let us therefore elaborate a little the problems which
arise there. The constituents of strongly interacting matter, the
hadrons, are finite size color-neutral bound states of two (mesons) or
three (nucleons) colored quarks coupled by colored gluons. At high
density, when the hadrons strongly overlap, each quark finds in its
immediate vicinity many others, so that a partitioning into subsets of
two or three quarks in a given volume no longer makes any sense,
suggesting that we now have quark matter. High temperatures lead
through particle-antiparticle production to high density. Hence we
expect at sufficiently high temperatures a transition from hadronic
matter, with color-neutral hadrons as constituents, to a plasma of
unbound colored quarks and gluons. One of the basic issues of strong
interaction thermodynamics is this quark-hadron transition.

\begin{figure}[htb]
\mbox{
\hskip0.5cm\epsfig{file=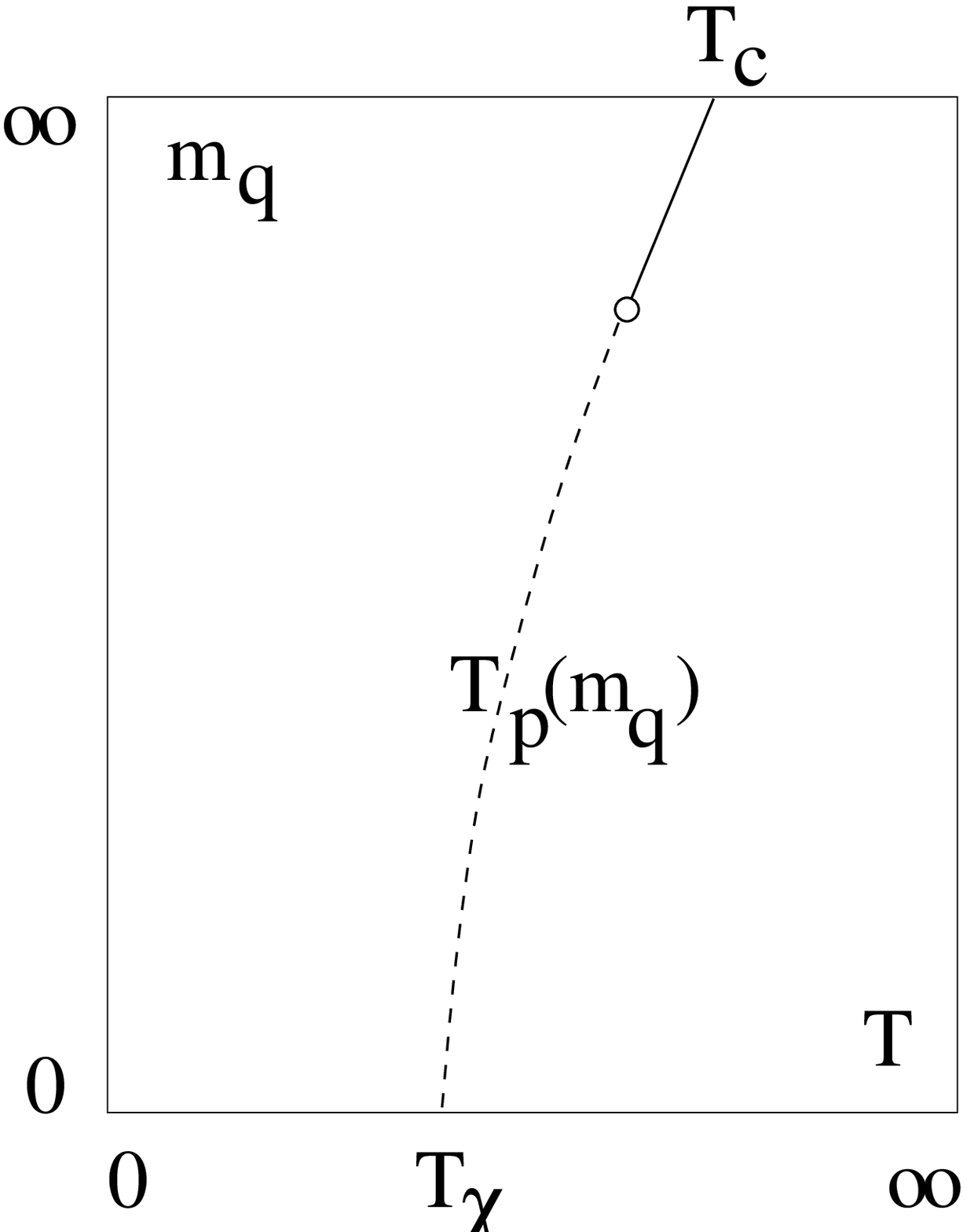,width=6cm,height=5.3cm}
\hskip2cm
\epsfig{file=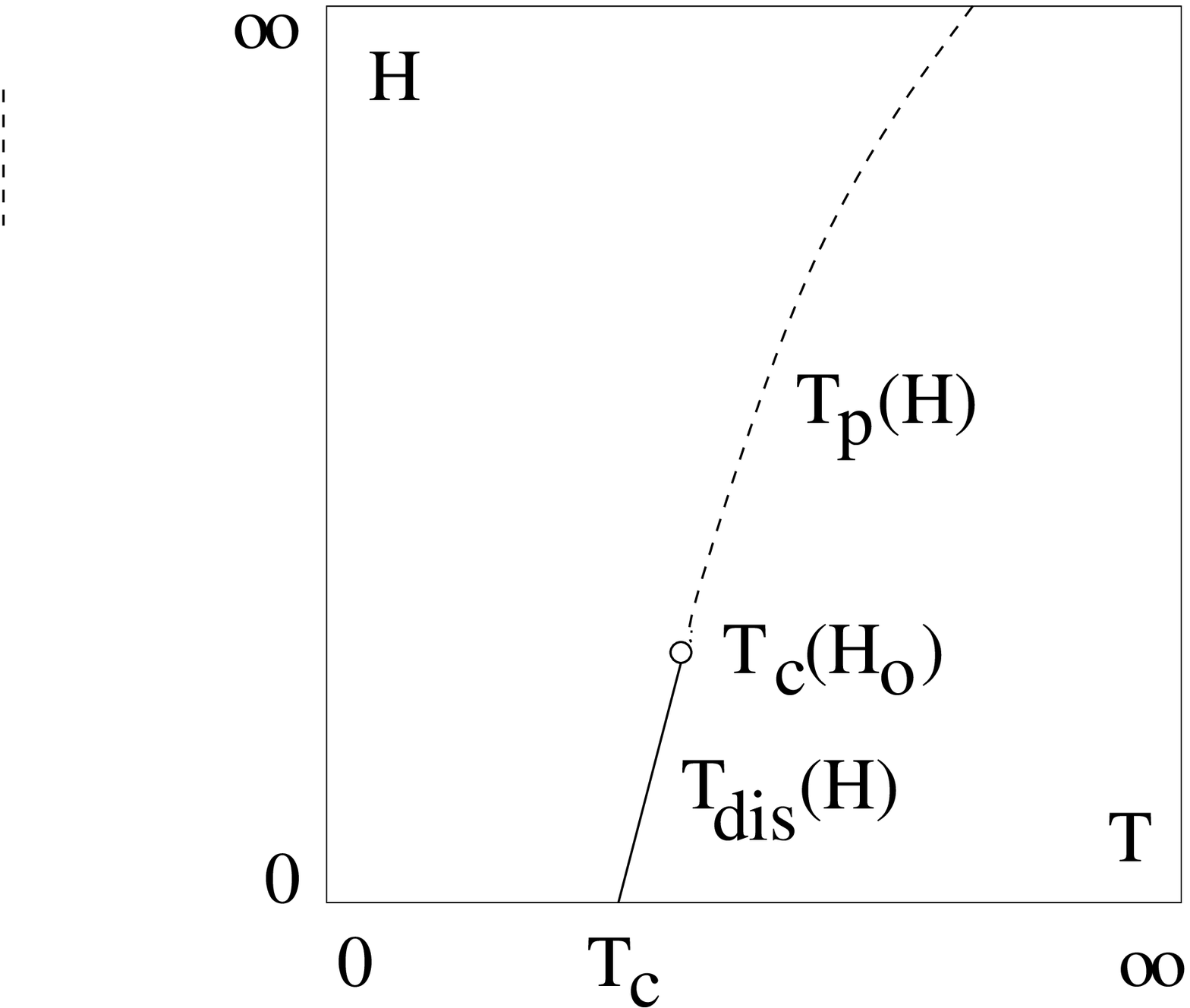,width=6cm, height=5cm}}
\vskip0.5cm
\caption{The phase structure of QCD (left) and of the 3-d 3-state 
Potts model (right)}
\label{3-1}
\end{figure}

The theory of strong interactions, quantum chromodynamics (QCD), is
defined by the QCD Lagrangian ${\cal L}(m_q)$, which contains the
bare quark mass as an open parameter. In QCD thermodynamics, two limits
are of particular interest.
For massless quarks, $m_q=0$, ${\cal L}(0)$ is invariant under chiral
transformations. This chiral symmetry is spontaneously broken for
$T < T_{\chi}$: through a surrounding gluon cloud, the quarks acquire a
dynamically generated `constituent quark' mass $M_q$, which essentially
determines the mass of the conventional hadrons. Thus the mass of a
nucleon as three quark bound state is approximately $3M_q$. At
$T=T_{\chi}$, chiral symmetry is restored, $M_q \to 0$ and hadrons
dissolve into quarks and gluons.
In the opposite limit, for $m_q \to \infty$, the quarks drop out and QCD
becomes a (non-Abelian) SU(3) gauge theory, which describes a system of
interacting gluons. At low temperatures, these bind to form `glueballs'
as the hadrons of the theory. The resulting Lagrangian has a global
$Z_3$ symmetry. For some $T \geq T_c$, this symmetry is spontaneously
broken; color screening causes the colorless glueballs to dissolve into
colored gluons, corresponding to the onset of color deconfinement.
QCD thus leads to two genuine phase transitions:
\begin{itemize}
\vspace*{-0.25cm}
\item{for $m_q=0$, chiral symmetry restoration at $T_{\chi}$,
corresponding to the vanishing of the constituent quark mass;}
\vspace*{-0.25cm}
\item{for $m_q=\infty$, spontaneous $Z_3$ breaking at $T_c$
corresponding to the onset of color deconfinement in pure gauge theory.}
\vspace*{-0.25cm}
\end{itemize}
In the real world, $m_q$ is small but finite; what happens then?
For three massless quark flavors, the $Z_3$ transition is first order,
so a discontinuity will persist for some range of masses 
$m_q^{\rm end} < m_q$. For $m_q < m_q^{\rm end}$, there is no
further thermal critical behavior, which 
means that there is no true phase transition separating hadronic matter
and quark-gluon plasma (see Fig.\ \ref{3-1}). Is the transition from one
state to the other then just a rapid cross-over? We would like to
speculate that it is the Kert\'esz line of QCD, corresponding to the
onset of percolation for clusters of deconfined matter \cite{HS}, and we 
have carried out some preliminary studies to address this idea.

\medskip

\noindent{\bf 2.\ Percolation and Deconfinement}

\medskip

As a simplified model, we consider here pure SU(2) gauge theory, which
describes a system of interacting gluons of three different colors.
Given such a medium at temperature $T$, we denote by $V(r,T)$ the
potential between two static color charges $Q$ and $\bar Q$ separated
by a distance $r$. Below a certain critical temperature $T_c$, there is
confinement: the potential rises linearly with distance, so that
$V(r,T) \to \infty$ for $r \to \infty$. Above $T_c$, color screening
limits the range of the potential, so that now $V(r,T) \to g(T)$ for $r
\to \infty$, with a finite $g(T) > 0$. The Polyakov loop expectation
value, defined by
\be
L(T) = \lim_{r \to \infty} \exp \{-V(r,T)/T\},
\label{2.1}
\ee
thus vanishes for $T \leq T_c$, i.e., in the confinement region,
and takes on a non-zero value for $T > T_c$, when deconfinement set in.
Hence it is the deconfinement order parameter.

In the lattice formulation of finite temperature QCD, one studies
systems in the usual three space dimensions, with one additional
dimension specifying the temperature. A Polyakov loop becomes in
lattice SU(2) an up or down spin of arbitrary amplitude, associated to
a given spatial lattice site and oriented in the temperature direction;
the expectation value $L(T)$ is the average over all spatial lattice
sites. Polyakov loop configurations on a given spatial lattice thus form
something like a continuous spin Ising model, and it was therefore
conjectured that the deconfinement transition in SU(2) gauge theory is
in the universality class of the Ising model \cite{SY}. This conjecture
was subsequently confirmed by extensive lattice studies, showing that
the critical exponents of the two models agree with remarkable precision
\cite{Engels}.

We now would like to check if the deconfinement transition can also be
described as percolation of like-sign Polyakov loop clusters. The first
question to be addressed is the generalization of the global FK bond
weight $1-\exp\{-2J/T\}$ of the conventional Ising model with $s_i=\pm
1$ to the case where the amplitude of the spin becomes a local variable.
It was shown in both analytic and numerical studies of the continuous
spin Ising model \cite{blanchard} that percolation and the
thermal transition in fact become equivalent if the bond weights are
taken to have the local form $1 - \exp\{-(2J/T)s_is_j\}$ for adjacent
like-sign spins.

Turning to SU(2) gauge theory, we therefore start with the local FK bond
weight $p(i,j) = 1 - \exp\{2 \kappa L_iL_j\}$ expressed in terms of the
local values $L_i,L_j$ of like-sign Polyakov loops on adjacent
spatial lattice sites $i,j$; the variable $\kappa$ must take into
account the temperature dependent Polyakov loop coupling. The
generalization to SU(2) thus leads to two basic problems: how can one
determine the functional form of $\kappa(T)$, and is it justified to
approximate the correlations provided by the theory to only nearest
neighbor interactions? In the strong coupling limit \cite{KG}, one finds
for $N_{\tau}=2$ that
$\kappa(T) \simeq (\beta(T)/2)^2$, where $\beta(T)$ is the known
temperature dependent coupling in the SU(2) gauge theory action. Using
this relation, we thus generate SU(2) configurations at different
temperatures and study the clustering properties. The results \cite{FS1}
show overall agreement, but some discrepancies in details; in
particular, the critical couplings for the thermal transition and for
deconfinement are obtained with very high precision and disagree
slightly.

It therefore seems worthwhile to consider a formulation in which the
strong coupling approximation and the restriction to nearest neighbor
interactions are removed. This can be done in an effective spin
description, in which the thermal system is parametrized in a spin model
allowing longer range interactions as well, with up to 20 different
couplings between different spins \cite{FKPS}. The values of these
couplings are then determined numerically near the thermal critical
point which in turn provides generalized bond weights $\kappa_n$,
\be
p(i,i+1)=1-\exp\{-2\kappa_1 L_iL_{i+1}\},~~
p(i,i+2)=1-\exp\{-2\kappa_1 L_iL_{i+2}\},~~...
\label{2.2}
\ee
The resulting agreement between the critical thermal and percolation
parameters is excellent, as seen in Table 1. The approach used here does
not solve the question of how to define in general the FK bond weights
in terms of the SU(2) Lagrangian; but it does show that the equivalence
between thermal and geometric descriptions can be established
numerically.

\begin{table}[htb]
\label{table:I}
\hskip 0.5cm
\newcommand{\m}{\hphantom{$-$}}
\newcommand{\cc}[1]{\multicolumn{1}{c}{#1}}
\renewcommand{\tabcolsep}{1pc} 
\renewcommand{\arraystretch}{1.2} 
\begin{tabular}{@{}lllll}
\hline
$$            & \cc{Critical point} & \cc{${\beta}/{\nu}$} &
\cc{${\gamma}/{\nu}$} & \cc{$\nu$} \\
\hline
Percolation                 & ~~1.8747(2)
&0.528(15) 
&1.985(13)  & 0.632(11) \\
Spontaneous $Z_2$ Breaking     & ~~1.8735(4)
&0.523(12)  
& 1.953(18)& 0.630(14) \\
Ising Model \cite{ferr} &   & 0.518(7)  &
1.970(11)  
& 0.6289(8)  \\
\hline
\end{tabular}\\[2pt]
\caption{Parameters for percolation and spontaneous $Z_2$ breaking in
3+1 SU(2).}
\end{table}

We therefore conclude that color deconfinement in SU(2) gauge theory can
be considered either as spontaneous $Z_2$ symmetry breaking or as the
onset of Polyakov loop percolation. Both phenomena occur at the same
critical temperature $T_c$ and are described in terms of the same set of
critical exponents.

\medskip

\noindent{\bf 3.\ Percolation in Non-Critical Thermal Systems}

\medskip

We would like to formulate a percolation description of deconfinement
in full QCD with dynamical quarks; again we start with a preliminary
study and consider spin systems with a non-vanishing external field $H
\not= 0$. The basic issue here is to generalize the FK bond weights
to include a dependence on $H$: $p_b(T,H)$. This problem is not yet
solved; we shall address two closely related questions.

Consider the 2-d Ising model with $H\not= 0$; the behavior of the
magnetization was shown in Fig.\ \ref{1-1}. Included there is the Kert\'esz
line $T_p(H)$, separating the percolating ($T \leq T_p(H)$) and the
non-percolating region ($T > T_p(H)$). In addition, one can also consider 
the pseudocritical line $T_{\chi}(H)$, defined by the temperature value 
for which the isothermal susceptibility $\chi_T \equiv (\partial m / 
\partial H)_T$ peaks for a given $H$; for $H=0$, it would diverge at $T_c$.

We want to compare these two temperature lines \cite{FS2}; in all
percolation calculations, we use the conventional ($H$-independent) FK bond
weights introduced in section 1. In terms of the reduced variables $t
\equiv (T-T_c)/T_c$ and $h \equiv H/J$ we find for small $h$
\be
t_{\chi}=c_1 h^{1/\beta \delta} ~~~{\rm and}~~~ t_p=c_2 h^x,
\label{3.1}
\ee
for the pseudocritical Ising line and the Kert\'esz line, respectively,
with constants $c_1$ and $c_2$. For the Ising model, one has the exact
result $1/\beta \delta = 8/15 \simeq 0.533$, while a numerical study of
percolation gives $x=0.534(3)$. For small $h$, the two lines thus have
the same $h$-dependence. However, $c_1 \not= c_2$, so that the lines do
not coincide. For large $h$, this is obvious: $t_p(h) \to t_b$ for $h
\to \infty$, where $t_b$ is the bond percolation temperature, but
$t_{\chi} \to \infty$ for $h \to \infty$, although the susceptibility
peak becomes arbitrarily weak in this limit. This illustrates the
necessity to obtain field-dependent bond weights $p_b(T,H)$ for a
reasonable description at $H\not= 0$. We note here that a comparison
with the pseudocritical behavior of the isothermal susceptibility is not
a unique way to specify $p(T,H)$, since other susceptibilities or the
correlation length lead to slightly shifted peak temperatures.

In a second study \cite{FS3}, we consider the 3-d 3-state Potts model,
which shows a first order phase transition for $H=0$. When an external
field is turned on, the discontinuity in e.g.\ the magnetization
decreases but continues for $0 \leq H < H_0$, defining a line $T_{\rm
dis}(H)$. This ends for some value $H=H_0$ at $T_c(H_0)$ with a second
order phase transition which appears to be in the universality class of
the Ising model \cite{KSt}. For $H > H_0$, the partition function is
analytic and there is no more thermal critical behavior (see Fig.\
\ref{3-1}). Using the conventional FK bond weights $p(T)=1-\exp\{-2J/T\}$, we
now study the percolation behavior of the model. The result is
\begin{itemize}
\vspace*{-0.25cm}
\item{for $0 \leq H < H_0$, there is a discontinuous onset of
percolation at the same $T_{\rm dis}(H)$ at which there is discontinuous
thermal behavior;}
\vspace*{-0.25cm}
\item{at $H=H_0$, the percolation and the thermal
transition become continuous and occur at the same temperature
$T_c(H_0)$. In contrast to the Ising-like thermal behavior, the
percolation transition leads to exponents which are neither in the Ising
nor in the random percolation universality class. This is presumably
due to the infinite correlations at $T_c(H_0)$, similar to what is
found for pure site percolation in the 2-d Ising model at the Curie
point \cite{SG}.}
\vspace*{-0.25cm}
\end{itemize}
Thus again there seems to exist a definite relation between percolation
and thermal critical behavior, but with the conventional FK weight,
there is not real equivalence; a bond weight $p(T,H)$ seems to be
needed. Here, however, in contrast to the pseudocritical line(s), the
problem is well-defined: what $p(T,H)$ will put percolation at
$T_c(H_0)$ into the Ising universality class? This is quite similar to
the original question of what $p(T)$ would change random percolation
enough to map the new clustering pattern onto thermal critical behavior,
solved by the FK bond weights. If a correct $p(T,H)$ can be determined,
one would have full equivalence between percolation and thermal critical
behavior both in the first order region and for the continuous
transition at the endpoint; beyond that, percolation only would persist.  
Given such a framework, one could then try to address
the similar structure in full QCD (Fig.\ \ref{3-1}).

\medskip

\noindent{\bf 4.\ Outlook}

\medskip

From what we have seen, thermal critical behavior can in many cases be
formulated as cluster percolation. The interesting feature is, however,
that percolation can persist even when thermal critical behavior stops,
such as in spin systems with external field. Here the percolation
strength $P(T)$ and the cluster size $S(T)$ remain singular when the
partition function $Z(T)$ becomes analytic. This raises some intriguing
questions. Can one give a more general definition of critical behavior,
such that percolation is included? What physical features distinguish
the percolating from the non-percolating phase \cite{adler}? Is it
possible to define a more `detailed' partition function $Z(T,x)$ at
some fixed connectivity measure $x$, which results in the usual $Z(T)$
after summation or integration over all $x$? It seems that all these
points could eventually provide a new opening to thermodynamics.

\medskip

\centerline{\bf Acknowledgements}

\medskip

Most of the recent results reported on here are based on joint work
with Santo Fortunato; it is a great pleasure to thank him for this
fruitful collaboration. Thanks go also to Ph.\ Blanchard, A.\ Coniglio,
D.\ Gandolfo, J.\ Kert\'esz and D.\ Stauffer for helpful discussions.
The financial support of the TMR network ERBFMRX-CT-970122 and the DFG
Forschergruppe Ka 1198/4-1 is gratefully acknowledged.



\begin{thebibliography}{99}

\bibitem{FK} C.\ M.\ Fortuin, P.\ W.\ Kasteleyn
J.\ Phys.\ Soc.\ Japan {26} (Suppl.), 11 (1969);\\
Physica {57}, 536 (1972).

\bibitem{CK} A.\ Coniglio, W.\ Klein,
J.\ Phys.\ A {13}, 2775 (1980).

\bibitem{K} J.\ Kert\'esz, Physica A 161, 58 (1989).

\bibitem{SY} B.\ Svetitsky, L.\ G.\ Yaffe, Phys. Rev. D {26}, 963 (1982).

\bibitem{Engels} J.\ Engels et al., \PL B 365, 219 (1996).

\bibitem{SW} R.\ H.\ Swendsen, J.\ S.\ Wang, Physica A {167}, 565
(1990).

\bibitem{HS} H.\ Satz, \NP A 642, 130c (1998).

\bibitem{FS1} S.\ Fortunato, H.\ Satz, \PL B {475}, 311 (2000);\\
\NP A 681, 466c (2001).

\bibitem{FKPS} S.\ Fortunato et al., \PL B {509}, 189
(2001).

\bibitem{blanchard} P.\ Bia{\l}as et al., \NP B 583, 368 (2000).

\bibitem{KG} F.\ Green, F.\ Karsch, \NP B 238, 297 (1984).

\bibitem{ferr} A.\ M.\ Ferrenberg, D.\ P.\ Landau, \PR B 44, 5081 (1991)

\bibitem{FS2} S.\ Fortunato, H.\ Satz, \PL B {509}, 189 (2001).

\bibitem{FS3} S.\ Fortunato, H.\ Satz, hep-ph/0108058.

\bibitem{KSt} F.\ Karsch and S.\ Stickan, \PL B 488, 319 (2000).

\bibitem{SG} M. F. Sykes, D. S. Gaunt, J. Phys. A {9},
  2131-2137 (1976).

\bibitem{adler} For first steps in this direction, see e.g.,
J. Adler, D. Stauffer, Physica A {175}, 222 (1991).

\end{thebibliography}
\end{document}